\documentclass[lettersize,journal]{IEEEtran}
\usepackage{amsmath,amssymb,amsfonts}
\usepackage{graphicx}
\usepackage{float}
\usepackage[caption=false,font=normalsize,labelfont=sf,textfont=sf]{subfig}
\usepackage{color}
\usepackage{cite}
\usepackage{multirow}
\usepackage{algorithm}  
\usepackage{algorithmicx}  
\usepackage{algpseudocode} 
\usepackage{bm}
\usepackage[colorlinks,linkcolor=black,anchorcolor=black,
citecolor=black,urlcolor=black]{hyperref}

\usepackage{stfloats}

\usepackage{siunitx}
\DeclareSIUnit\bps{bps}
\DeclareSIUnit\Torr{Torr}
\DeclareSIUnit\torr{Torr}
\DeclareSIUnit\sample{Sa}
\newcommand*{\circled}[1]{\lower.7ex\hbox{\tikz\draw (0pt, 0pt)%
  circle (.5em) node {\makebox[1em][c]{\small #1}};}}

\graphicspath{{figures}}
\begin{document}

 \title{Still Waters Run Deep: Extend THz Coverage with Non-Intelligent Reflecting Surface}

\author{Chong Han,~\IEEEmembership{Member,~IEEE,}, Yuanbo Li, Yiqin Wang
\thanks{Chong Han, Yuanbo Li, and Yiqin Wang are with the Terahertz Wireless Communications (TWC) Laboratory, Shanghai Jiao Tong University, Shanghai, China (e-mail: \{chong.han, yuanbo.li, wangyiqin\}@sjtu.edu.cn).}}

	{}
	\maketitle
	\thispagestyle{empty}


\begin{abstract}
Large reflection and diffraction losses in the Terahertz (THz) band give rise to degraded coverage abilities in non-line-of-sight (NLoS) areas. To overcome this, a non-intelligent reflecting surface (NIRS) can be used, which is essentially a rough surface made by metal materials. NIRS is not only able to enhance received power in large NLoS areas through rich reflections and scattering, but also costless and super-easy to fabricate and implement. In this article, we first thoroughly compare NIRS with the lively discussed intelligent reflecting surface (IRS) and point out the unique advantages of NIRS over IRS. Furthermore, experimental results are elaborated to show the effectiveness of NIRS in improving coverage. Last but not least, open problems and future directions are highlighted to inspire future research efforts on NIRS.
\end{abstract}


\section{Introduction}
\IEEEPARstart{O}{ver} the past few decades, wireless communication networks have experienced revolutionary developments, from the first generation (1G) to the most recent fifth generation (5G). 
Nonetheless, looking towards 2030, the mobile communication network will further evolve to the sixth generation (6G), where internet-of-everything (IoE) is expected to be achieved with ubiquitous network coverage and massive connectivity~\cite{Giordani2020Toward}. Various and abundant intelligent devices, such as smartphones, mixed reality (MR) headsets, as well as sensors and machines, will generate a large amount of message and data for wireless communications. As electromagnetic infrastructure to support them, ultra high data rates (e.g., up to 1 Terabits per second) are needed, which can not be fullfilled by current spectrum resource and thus motivate the exploration of the Terahertz (THz) band. Spanning the frequency between \SI{0.1}{THz} to \SI{10}{THz}, the THz band is envisioned as a key technology to address the spectrum scarcity and capacity limitations of current wireless systems~\cite{Akyildiz2022THz,9583918}, thanks to its broad contiguous bandwidth (from tens up to hundreds of GHz). 
\par Wonderful as THz communication is, however, it has its own drawbacks. Among others, one key problem is the weak coverage ability of THz communications in non-line-of-sight (NLoS) areas. At high frequencies, reflection, diffraction, and penetration losses worsen~\cite{han2015multiray,Jacob2012Diffraction,eckhardt2021channel}. As a result, when line-of-sight (LoS) transmission is blocked, sometimes drastic degradation of link quality may occur. To address LoS blockage problem, one natural solution is to add more active nodes in the network, such as base stations (BSs), access points (APs), and active relays, which however, associate with extra hardware and energy costs. By contrast, energy efficient solutions are preferred, such as intelligent reflecting surface (IRS), which is a passive tunable metasurface and able to redirect propagating THz waves~\cite{Bjornson2020Reconfigurable}. 
However, even though the theoretical performance of IRS is extraordinary, realization of IRS in the THz band might be difficult and far from practice, for the following reasons. First, due to high frequencies of THz waves, thousands of elements are required to compensate for the large path loss from IRS to receiver. The fabrication of such large number of IRS elements and corresponding control circuits might be very difficult and costly. Second, attributed to the small wavelength of THz waves, tiny antenna elements in the order of sub-millimeter could be fabricated and densely placed to form ultra massive multiple-input-multiple-output (UM-MIMO) system, resulting in improved spectral efficiency and coverage capability. Integrating UM-MIMO and IRS, the concatenated channel from transmitter (Tx) to IRS to receiver (Rx) is expressed with a channel tensor with dimensions of $N_t\times N_{\text{IRS}}\times N_r$, with $N_t$, $N_\text{IRS}$, and $N_r$ denoting the numbers of elements of transmitter array, IRS, and receiver array, respectively. The massive IRS elements would make the accurate channel estimation of the large-scale channel tensor computationally complex, for which the joint optimization of the UM-MIMO and IRS is hard to achieve. Therefore, still a long and spiny path is in front to practically implement IRS in the THz band.
\par By contrast, a more realistic and easier way is to use non-intelligent reflecting surface (NIRS), which is essentially a rough surface simply made of metal materials. Compared to IRS, NIRS loses the ability to adapt to mobile users or suppress interference from neighbouring BS, while gaining advantages such as nearly no cost, no fabrication, and super-easy deployment. We hereby note that NIRS is different from the frequently mentioned reflectors in cmWave and mmWave bands~\cite{Khawaja2020Coverage} as follows. NIRS is rough and require no specific design, while reflector is a smooth surface acting as electromagnetic mirrors. The reason that NIRS is preferred than reflectors is two-fold. On one hand, due to the small wavelength, THz waves are more sensitive to surface roughness, resulting in a stricter requirement for a reflector to be smooth considering the sub-millimeter wavelength. Lower fabrication difficulty is the key advantage of NIRS, compared to reflectors. On the other hand, the high sensitivity of THz waves lead to strong scattering, i.e., non-specular reflections, especially when interacting with rough metal surfaces. Even though NIRS performs worse than reflectors in specular directions, it can simultaneously enhance signal strength in non-specular directions, thus covering wider NLoS areas. In summary, even though NIRS appears more clumsy compared to IRS or reflectors, the outstanding low cost, low utilization difficulty, and wide coverage of NIRS promote it to be a good technique  for coverage extension in THz networks.

\begin{figure*}[!tbp]
    \centering
    \subfloat[]{ 
    \includegraphics[width=1.0\columnwidth]{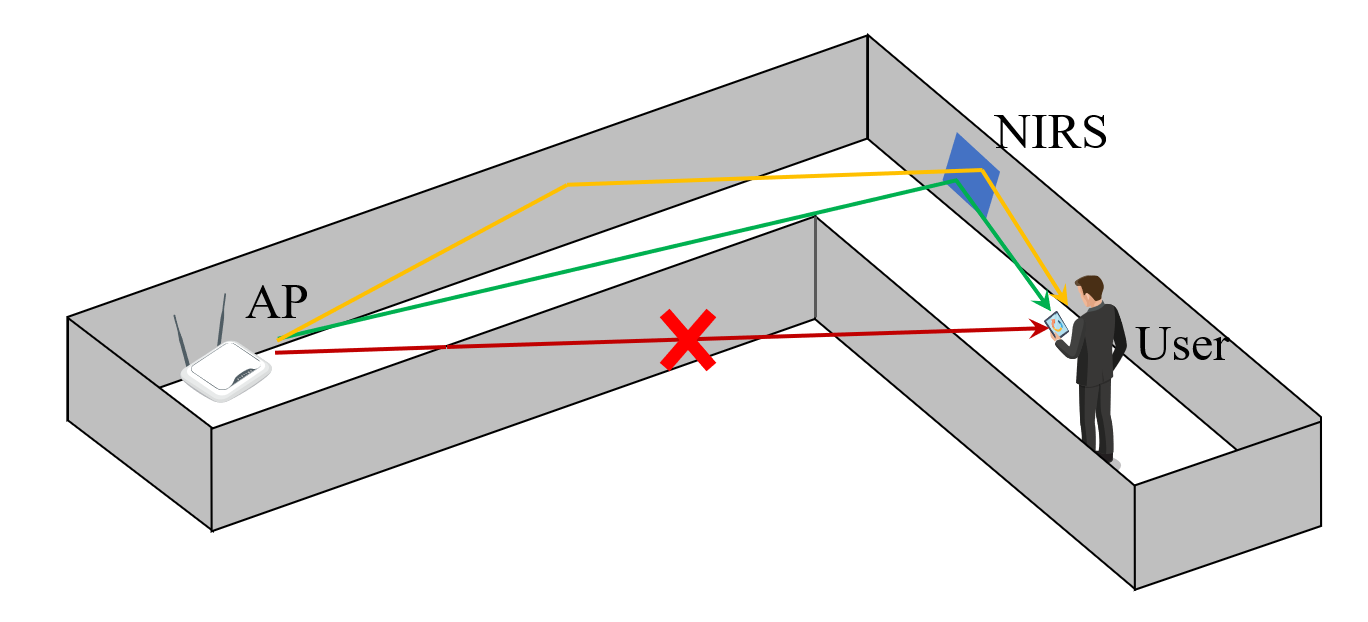}
    }
    \subfloat[]{ 
    \includegraphics[width=0.95\columnwidth]{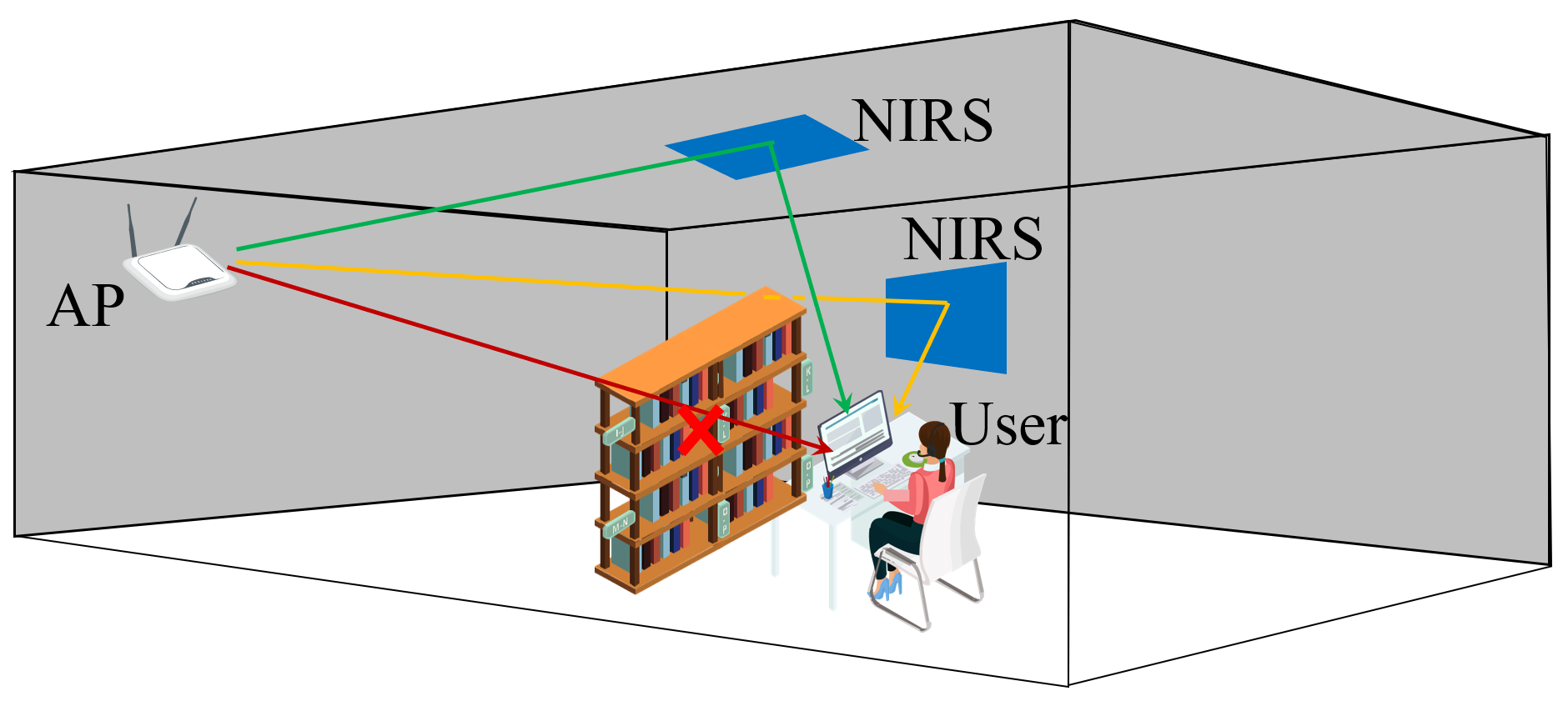}  
    }   
    \quad
    \subfloat[]{ 
    \includegraphics[width=0.9\columnwidth]{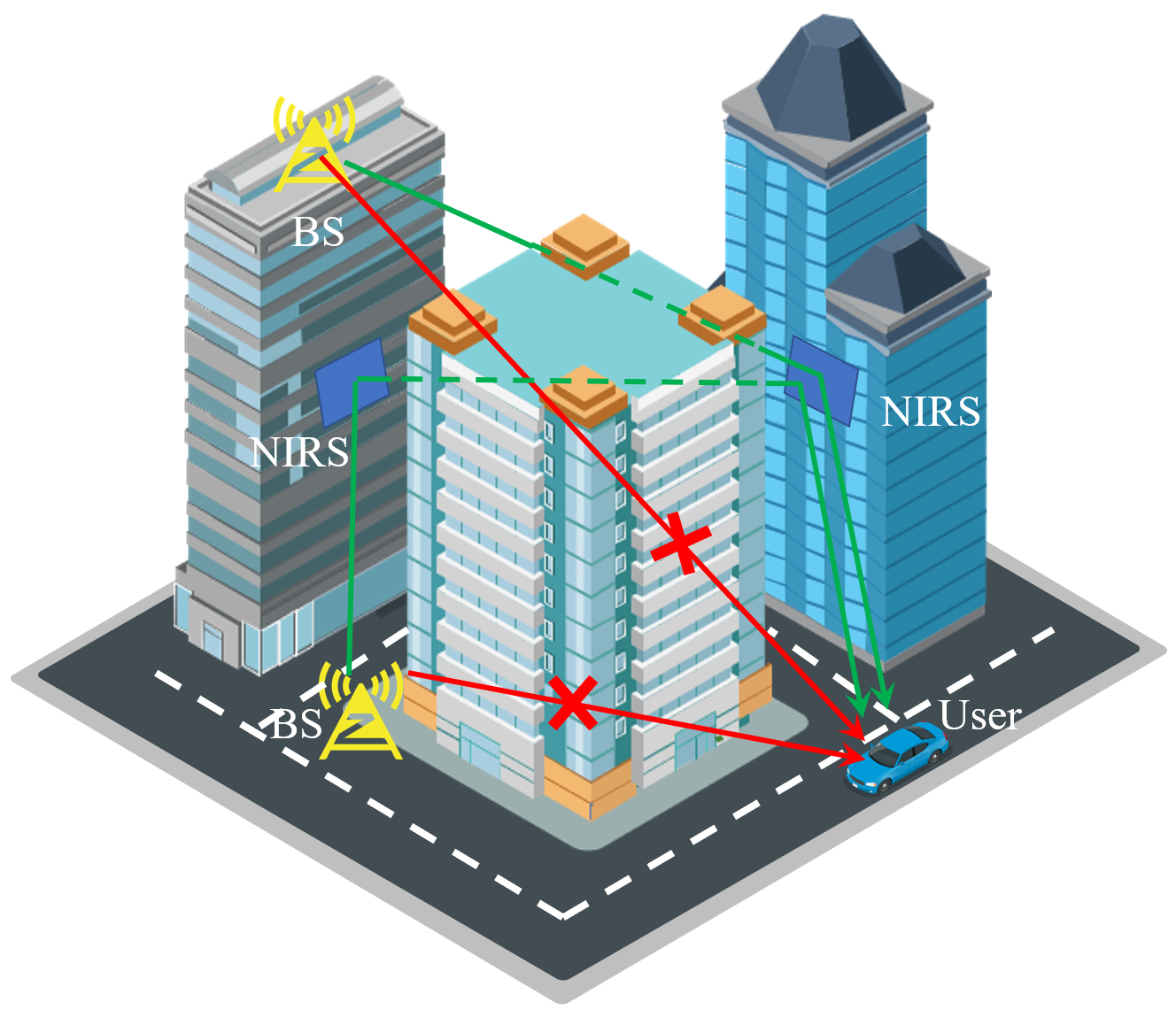}  
    }
    \subfloat[]{ 
    \includegraphics[width=1.0\columnwidth]{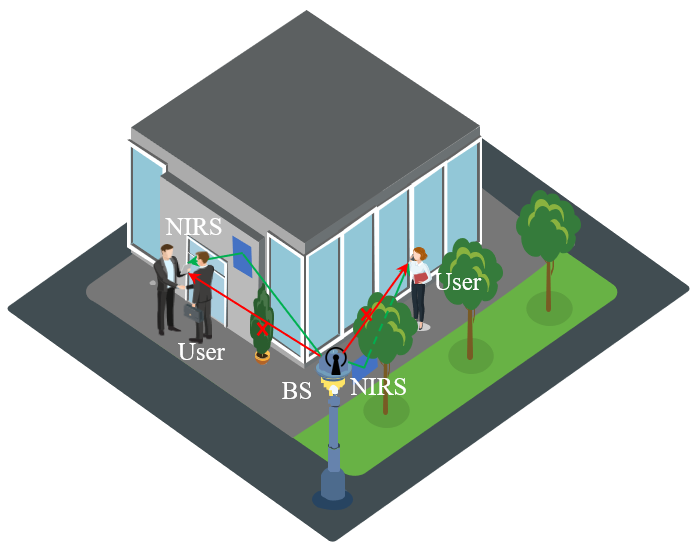}  
    }
    \caption{Illustration of use cases of NIRS, to overcome blockage of (a) walls in corridors, (b) objects in indoor rooms, (b) buildings in urban, (d) human or foliage.}
    \label{fig:illu}
    \vspace{-0.5cm}
\end{figure*}
\par Fig.~\ref{fig:illu} illustrates several typical use cases of NIRS, in both indoor and outdoor scenarios. For instance, in L-shaped corridors, the LoS path is blocked by walls, which results in significant link performance degradation. In light of this, NIRS can be deployed on the walls near the turning corner, to provide once-scattering or high-order reflections path to enhance the coverage in the NLoS region. Moreover, objects in indoor rooms, such as bookshelf in library, server rack in data center, etc., can also shelter the receivers from the access points. In this regard, the NIRS can be deployed on walls or ceilings, to bypass the blocking objects. Similarly, the blockage of high-rise buildings in urban areas can also be address by deploying NIRS on building surfaces. Last but not least, for pedestrian communicating with lamppost base stations, the blockage of human body and foliage can be severe, due to the weak penetration ability of THz waves. Therefore, the NIRS deployed on nearby walls or grounds can help redirect a reliable link.
\par To this end, NIRS is a promising technique for THz communications to address LoS blockage problem, by exploiting the benefits from rough surface scattering. Motivated by this, we provide an overview of NIRS, including the main advantages and disadvantages of NIRS compared to IRS, in terms of flexibility, fabrication and design difficulty, and compatibility to UM-MIMO systems. Moreover, to show the efficacy of NIRS, a preliminary experiment in an indoor corridor scenario is elaborated. Coverage and capacity are improved with deployment of NIRS. 
Furthermore, open problems and future challenges are highlighted to inspire future research, including NIRS channel modeling, reliable design, deployment and coordination optimization, joint communication and sensing enhancement.
\section{IRS v.s. NIRS: Trade-off of Performance and Implementation Difficulty}
\begin{table}[]
    \centering
    \caption{{Comparison of IRS and NIRS. For path loss comparison, $d_1$ and $d_2$ denotes the distance from Tx to IRS and from IRS to Rx, respectively. $n$ stands for the path loss exponent.}}
    \includegraphics[width=0.9\columnwidth]{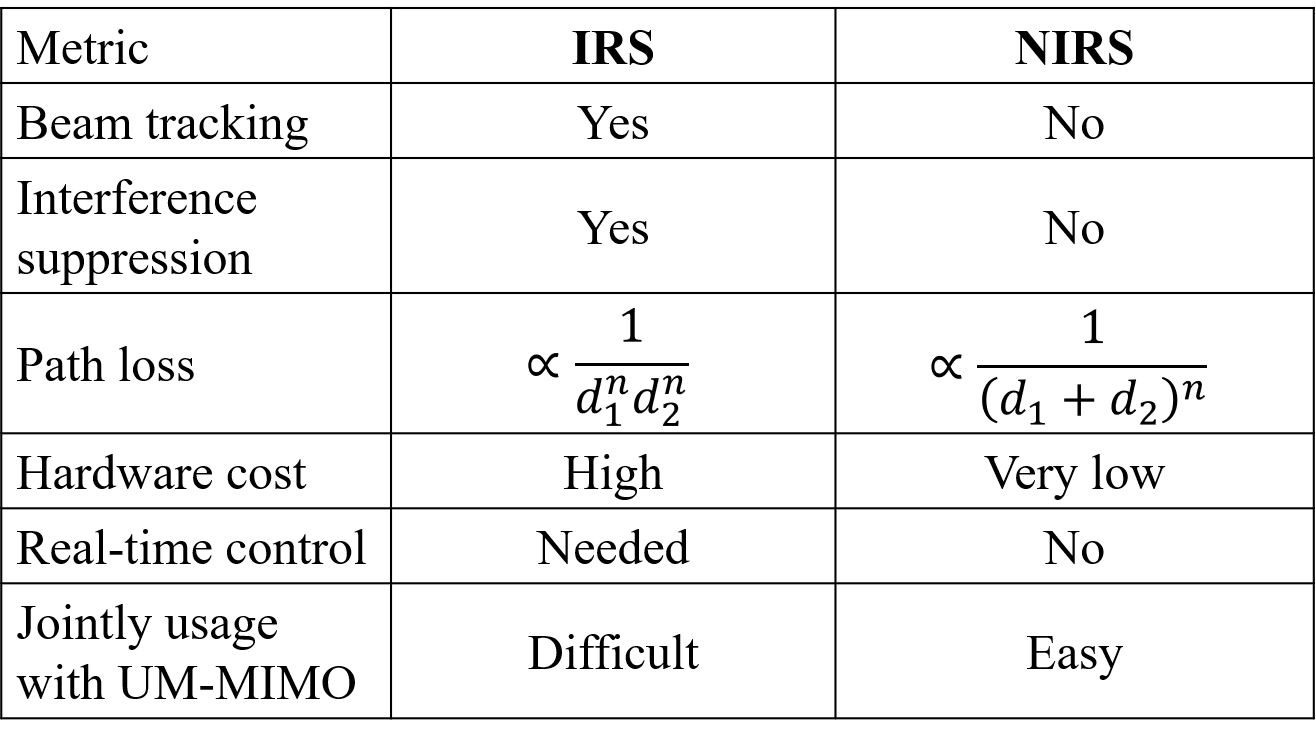}
    \label{tab:comparison}
    \vspace{-0.5cm}
\end{table}
\par As an overview, the comparison of IRS and NIRS is shown in Table~\ref{tab:comparison}. In short, the attractive excellent flexibility and intelligence of beam control via IRS comes at prices of high hardware and computation costs, which may prevent its realization in the THz band. On the contrary, the design and usage of NIRS are much easier in practice, with noticeable coverage enhancement. 
\subsection{Beam Control Ability}
\par High flexibility and adaptability are the key advantage of IRS compared to NIRS, which also distinguish their intelligence. IRS is composed of a large number of reflecting elements, which can manipulate the reflecting amplitude and phase shift of the impinging THz waves. Enabled by them, IRS has the so-called passive beamforming ability, i.e., the scattering pattern can be electrically controlled to realize certain purposes. For instance, when used for coverage extension, IRS can intelligently concentrate the signal power towards the directions of users. As the user moves, the IRS beam constantly steers to follow, achieving reliable communication links. Moreover, just like active beamforming of antenna arrays, IRS can also create zero-directions to suppress interference from neighbouring BS in down-link or to them in up-link. 

In this regard, NIRS is much more clumsy and thus non-intelligent. It can neither simultaneously change the scattering pattern to track mobile users, nor cancel interference in multi-user or multi-BS scenarios. In fact, NIRS may cause more interference since the NLoS signals are all enhanced, no matter from the targeted BS/AP or other BSs/APs. Therefore, before deployment in realistic communication networks, it is of importance to appropriately design and deploy the NIRS in positions maximizing the signal-to-interference and noise ratio (SNR). 
\subsection{Link Path Loss}
\par A key factor to evaluate link performance is the path loss, which determines the SNR at the receiver side. Due to the different design considerations, the path losses of NIRS and IRS-aided communication links are fundamentally different. Particularly, for IRS, the communication channel is concatenated by two segments, namely the Tx-IRS channel and the IRS-Rx channel. Moreover, to steer the IRS beams toward any user location, IRS elements are designed to scatter the incident signal omnidirectionally. In other words, the reflected signal from a single IRS element is just like being radiated by an omnidirectional antenna. As a result, the path loss of IRS-aided communication link is inversely proportional to the product of distance from Tx to IRS and distance from IRS to Rx, as clearly explained as the \textit{product-distance path loss model} in~\cite{9326394}. Considering only the LoS path, by using the Friis' formula, the free space path loss at \SI{300}{GHz} could exceed \SI{80}{dB} for a distance of only \SI{1}{m}. As a result, an overall path loss of Tx-IRS-Rx link would surpass \SI{160}{dB} if Tx-IRS and IRS-Rx distances are both \SI{1}{m}. To overcome the severe path loss, high beamforming gains and thus very large amount of IRS elements are needed. For example, as analyzed in~\cite{Nguyen2022Channel}, to outperform a direct link, the number of IRS elements needs to exceed 4096 in the THz band.
\par Unlike the omnidirectional IRS elements that spread signal energy uniformly to all directions, the reflected/scattered signals from NIRS is concentrated on several directions, such as the specular direction. Therefore, the path loss of Tx-NIRS-Rx link consists of two parts, namely the spreading loss part and additional reflection loss part. The spreading loss part is dependent on the overall link distance, i.e., reversely proportional to the summation of Tx-NIRS and NIRS-Rx distances, following the \textit{sum-distance path loss model}~\cite{9326394}. Moreover, the additional reflection loss part is dependent on reflection angles of NIRS-Rx direction. In strong reflection/scattering directions, such as the specular direction, the additional reflection loss could be as low as several dB, while in other directions, the additional reflection loss could increase up to tens of dB. Generally speaking, an overall path loss of \SIrange{100}{140}{dB} could occur for NIRS-aided links, depending on propagation distance and receiver locations~\cite{li2023channel}. Nonetheless, compared to the situations in NLoS areas without NIRS, the path loss values are mitigated by \SIrange{3}{17}{dB}.
\subsection{Fabrication and Control Difficulty}
\par The super-easy fabrication and control is the key strength of NIRS over IRS. As mentioned above, to explore the potential of IRS in the THz band, thousands of elements are needed to compensate for the significant path loss, each of which is electrically controlled. Moreover, due to the small wavelength of THz waves, IRS elements need to be rather small to omnidirectionally scatter the signals, e.g., a fifth of the wavelength (tends of micrometer). Such ultra-massive yet ultra-tiny IRS elements, and more importantly, their corresponding control circuits, would make it extremely costly and difficult to fabricate. On the contrary, NIRS require no specific design, whose fabrication is much easier and costless. As reported in~\cite{li2023channel}, NIRS can be simply made with super cheap aluminium foils, which yet realize considerable coverage extension for THz communications.
\subsection{Jointly Usage With UM-MIMO}
To exploit high spectral efficiency with spatial multiplexing, UM-MIMO is a critical mass in the THz band~\cite{9216613}. When embedding IRS in UM-MIMO systems, the joint optimization requires knowledge of the concatenated Tx array-IRS-Rx array channel, which however, is computationally complex due to the high dimensions. By contrast, the joint usage of NIRS and UM-MIMO system is natural since the NIRS needs no real-time control and thus does not add additional signal processing burden. Moreover, since NIRS creates rich reflection and scattering, the spatial degree of freedom improves, which is originally limited in the THz UM-MIMO system due to the sparse THz channel. Consequently, the rank of the UM-MIMO channel matrix and spatial multiplexing gain increase, which further improves the channel capacity.
\section{Experimental Results For NIRS-aided THz Coverage Extension}
\par In this section, we elaborate an experiment of using NIRS to extend THz coverage in a corridor scenario. The experiment set-up, deployment, and results are explained in the following part.
\subsection{Experiment Set-up and Deployment}
\par Experiments are conducted with a vector network analyzer (VNA)-based channel sounder, which is introduced in detail in~\cite{li2022channel}. Specifically, the measured frequency bands are \SIrange{306}{321}{GHz} and \SIrange{356}{371}{GHz}. During the channel measurement, the transmitter is only equipped with a standard waveguide WR2.8 for large coverage with wide beam, which has \SI{7}{dBi} antenna gain and a $30^\circ$ half-power beamwidth (HPBW). For the Rx side, to obtain omnidirectional channel observations, direction-scan sounding (DSS) scheme is utilized. Particularly, equipped with a directional horn antenna with a \SI{25}{dBi} antenna gain and a $8^\circ$ HPBW, the Rx scans the spatial domain with $10^\circ$ angle steps, from $0^\circ$ to $360^\circ$ in the azimuth plane and $-20^\circ$ to $20^\circ$ in the elevation plane.
\par
\begin{figure}[!tbp]
    \centering
    \includegraphics[width=0.76\columnwidth]{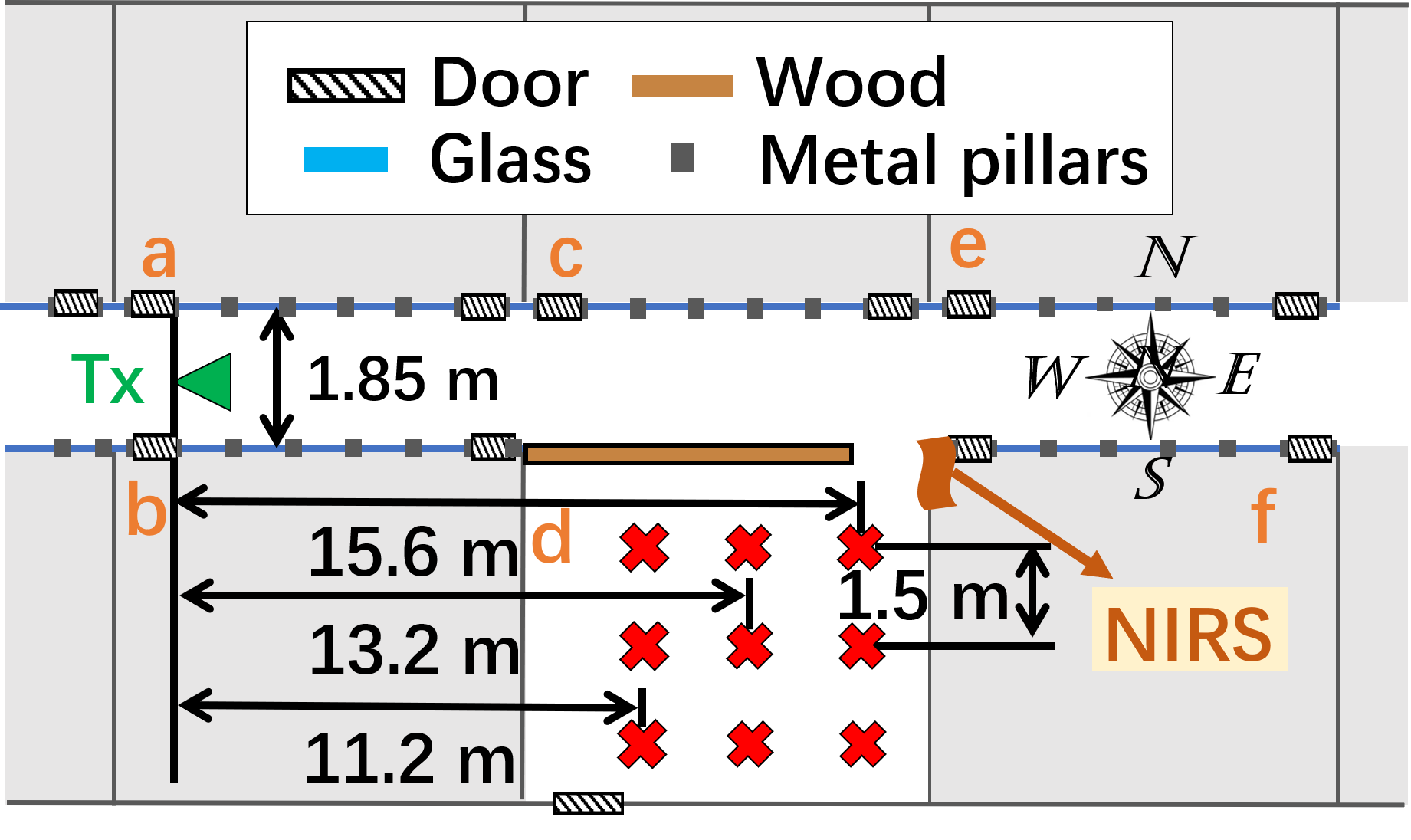}  
    \caption{The measurement deployment in the corridor.}
    \label{fig:deploy_corridor}     
    \vspace{-0.5cm}
\end{figure}

\begin{figure}[!tbp]
    \centering
    \includegraphics[width=0.76\columnwidth]{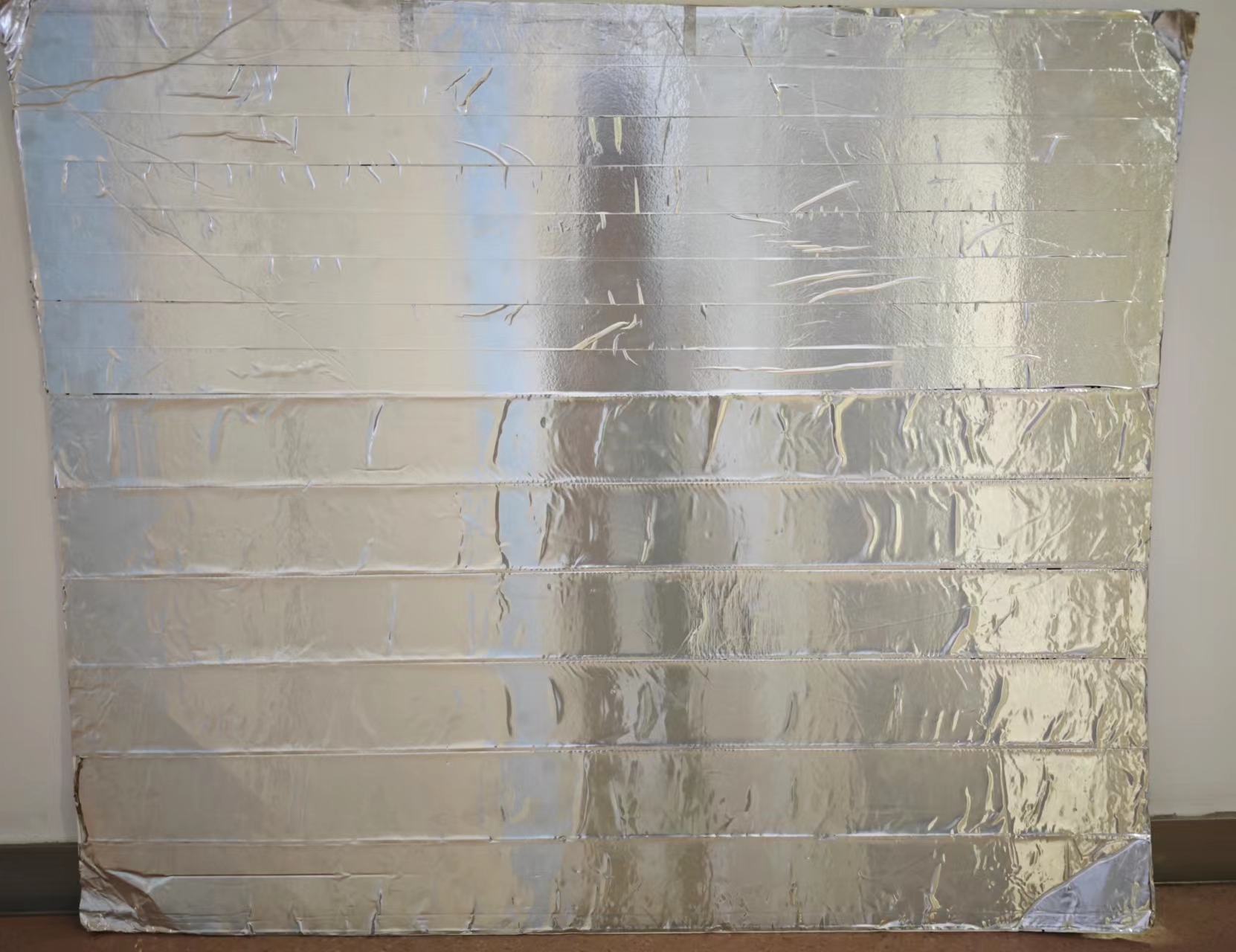}  
    \caption{The homemade NIRS. Aluminium foils are manually pasted on a foam board.}
    \label{fig:nirs}   
    \vspace{-0.5cm}
\end{figure}

 The measurement scenario is an indoor corridor on the second floor of the Longbin Building in Shanghai Jiao Tong University. Particularly, the transmitter is placed in the middle of the corridor near room a and remains static, while 9 Rx positions in NLoS areas in room d are selected, as shown in Fig,~\ref{fig:deploy_corridor}. To test the effectiveness of NIRS, a homemade NIRS with a size of \SI{1.2}{m}$\times$\SI{1.2}{m} is glued near the turning corner. As shown in Fig.~\ref{fig:nirs}, the NIRS is a foam board overlaid by aluminium foils that are manually pasted, resulting in a rough and irregular metal surface.
\subsection{Power Enhancement By Including NIRS}
\begin{figure}[!tbp]
    \centering
    \subfloat[]{ 
    \includegraphics[width=0.45\columnwidth]{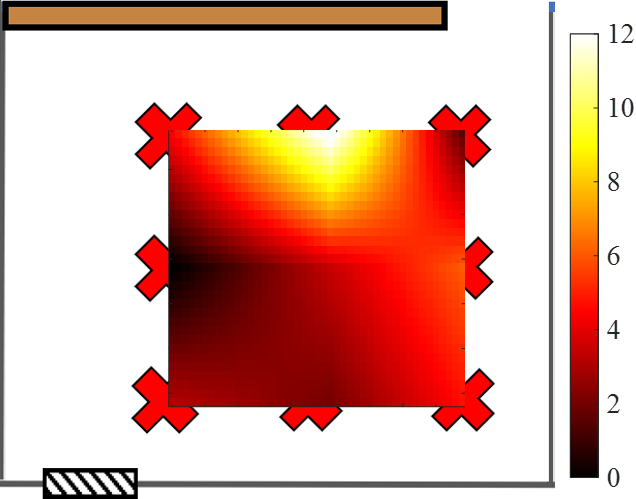}  
    }
    \subfloat[]{ 
    \includegraphics[width=0.45\columnwidth]{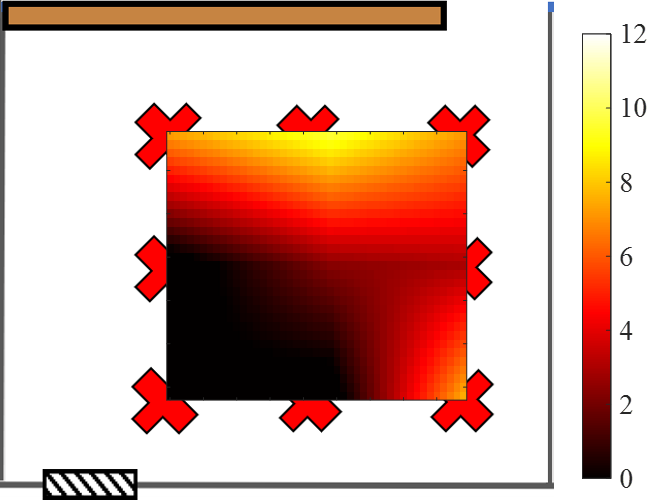}  
    }   
    \caption{The power enhancement by adding the NIRS in (a) 306-321 GHz, (b) 356-371 GHz.}
    \label{fig:cov}
    \vspace{-0.5cm}
\end{figure}
\par To observe the performance of NIRS, we compare the measured path loss with/without NIRS. Since only limited Rx positions are measured due to high time consumption of channel measurements, to analyze the coverage situations in the whole area, the path loss in positions between adjacent Rx locations are obtained through linear interpolation. As a result, the power enhancement is calculated as the difference of path loss before/after adding the NIRS, where the results are shown in Fig.~\ref{fig:cov} and several observations are made as follows. 
\par First, for both \SIrange{306}{321}{GHz} and \SIrange{356}{371}{GHz} bands, received power is enhanced in most areas. Specifically, 63.6$\%$ area at \SIrange{306}{321}{GHz} and 51.6$\%$ area at \SIrange{356}{371}{GHz} obtains power enhancement of more than \SI{3}{dB}. Moreover, the maximum power enhancement is \SI{12.56}{dB} and \SI{9.56}{dB} at \SIrange{306}{321}{GHz} and \SIrange{356}{371}{GHz}, respectively. This proves the effectiveness of the NIRS. Second, the power enhancement by adding the NIRS is not uniform in the NLoS areas. At certain Rx positions, such as the top-middle receiver location, the path loss is decreased by nearly \SI{10}{dB}, while in other Rx positions, especially those receiver positions that are far away from the NIRS, the received power barely changes. Third, comparing the two frequency bands, the power enhancement shows different patterns. Therefore, it might be hard to control the NIRS to enhance the received power at certain Rx locations.
\subsection{Channel Capacity With/Without NIRS}
\begin{figure}
    \centering
    \includegraphics[width=0.9\columnwidth]{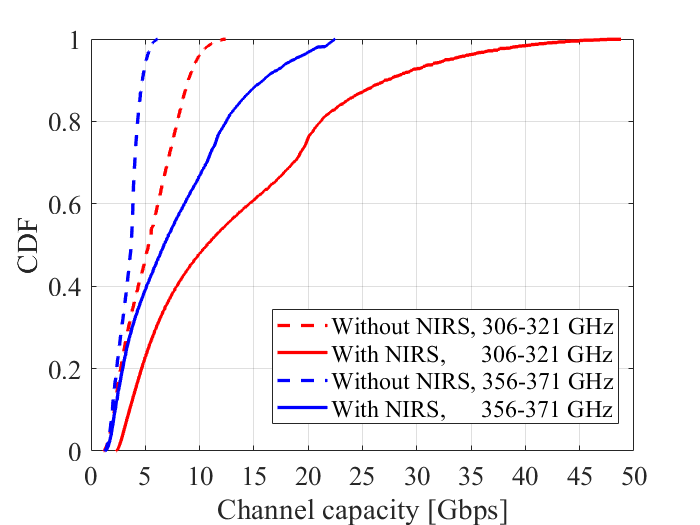}
    \caption{Channel capacity with/without NIRS.}
    \label{fig:cap}
    \vspace{-0.5cm}
\end{figure}
To clearly show the effectiveness of NIRS, SNR is calculated based on the measured path loss results, assuming a realistic THz communication link with reference parameters in~\cite{Rikkinen2020thz}. Key parameters include a bandwidth of \SI{15}{GHz}, transmitter power of \SI{13}{dBm}, Tx and Rx antenna gains of \SI{25}{dB}. Furthermore, based on the SNR results, the channel capacity can be evaluated, as shown in Fig.~\ref{fig:cap}. The results show that by adding the NIRS, the channel capacity in NLoS areas are greatly increased, especially in the \SIrange{306}{321}{GHz} band. Specifically, the average channel capacity increases from \SI{5.42}{Gbps} to \SI{13.55}{Gbps} at \SIrange{306}{321}{GHz}, and from \SI{3.46}{Gbps} to \SI{7.97}{Gbps} at \SIrange{356}{371}{GHz}, respectively. Moreover, with NIRS, in the best ten percent areas, the channel capacity exceeds \SI{27.08}{Gbps} and \SI{15.85}{Gbps} at \SIrange{306}{321}{GHz} and \SIrange{356}{371}{GHz}, respectively, while the values are only \SI{8.96}{Gbps} and \SI{4.73}{Gbps} at these two frequency bands without NIRS, respectively. Therefore, by including the NIRS, the channel capacity doubles or even triples than without using NIRS, proving its effectiveness. 
\section{Open Problems and Future Directions}
To effectively make use of NIRS, several open problems need to be addressed, including the channel modeling of the NIRS-aided communications, reliable design for site-specific coverage extension, optimal deployment and coordination of multiple NIRS, and possible joint communication and sensing enhancement. 
\subsection{NIRS Channel Modeling}
\par Unlike IRS and reflectors, which involve either diffusely scattering or specular reflection, the scattering phenomenon on rough NIRS depends on multiple factors, including surface roughness, material, surface size, etc. To characterize NIRS channels and further evaluate the link performance of NIRS-aided communications, an accurate yet efficient channel model is necessary. Since the problem of wave scattering from rough surfaces has no closed-form solutions, existing studies usually use approximate solutions. For instance, the Kirchhoff scattering theory might be used to calculate the scattering efficient, which further depends on rough surface height standard deviation~\cite{han2015multiray}. However, since the fabrication and design of NIRS is casual, such quantitative characteristics might not be available. Therefore, other models, such as statistical models or empirical fitting results based on real measurements, may be preferred in practice.
\par Another key problem related to NIRS channel modeling is the assessment of multipath richness and near-field effect resulted by NIRS. As mentioned above in Sec.II-D, NIRS can be easily embedded into UM-MIMO systems. The spatial multiplexing gain and channel capacity of UM-MIMO links highly rely on the number of significant paths in the communication channels. Therefore, by involving NIRS, the surrounding environment become more sensitive to THz waves, for which the originally weak high-order reflection/scattering paths can become more significant. Meanwhile, far-field propagation might convert to near-field, or cross near-and-far-field after experiencing NIRS scattering. Thus, the spatial multiplexing gain and channel capacity may increase. Extensive channel measurements are needed to analyze the multipath channel and analyze channel capacity of NIRS-aided THz communications.
\subsection{Reliable NIRS Design for Site-Specific Coverage Extension}
\par With the low fabrication cost, NIRS can enhance the THz coverage in the NLoS areas, as discussed in Sec. III. However, casual design of NIRS brings drawbacks such as random scattering pattern. Without careful design, it is hard to control which NLoS area to be enhanced, except for the specular directions that always receives better coverage due to smaller reflection loss. However, usage of NIRS in reality may be very site-specific, i.e., one usually has a target area or direction whose coverage needs to improved. In such cases, the random scattering pattern of NIRS prevents its effective usage. 
\par There are several possible research directions to address this issue. First, accurate modeling of the scattering pattern could enable reliable designs for practical use, which however, might be difficult due to the reasons aforementioned. Second, one way to obtain desired scattering pattern is to control the roughness of NIRS. For this purpose, special structure might be explored, such as placing polished metal cubes of different heights in a grid pattern. This scheme is potential, yet causing design difficulty. Third, a possible cost-effective and simple solution is to add more NIRS in the NLoS area, similar to the usage of double IRS for improved performance~\cite{Zheng2021Double}. However, this may incur interference problems, for which the joint deployment and coordination of multiple NIRS need to be considered.
\subsection{NIRS Deployment and Coordination Optimization}
\par Since NIRS is unable to change the beam steering direction after placement, where to deploy the NIRS is a key question to investigate. Generally speaking, since the specular reflection produces the strongest reflection, it is beneficial to place the NIRS in the specular reflection points between transmitter and receiver. Nonetheless, since NIRS can also enhance high-order reflections and scattering,  practical deployment is rather more complicated. Moreover, considering the mobility of users, it is usually preferred to obtain a good coverage enhancement in most of the NLoS area, rather than great improvement at several locations while neglecting others. Therefore, the NIRS deployment optimization is a question that needs to be answered.
\par Furthermore, as the NIRS scattering pattern is hard to control, it is intuitive to place more NIRS to fully cover the NLoS areas. Ideally speaking, by placing multiple NIRS in the appropriate positions, the coverage ability of THz communications can be greatly extended. The first NIRS closest to Tx can cover part of the NLoS area with first-order reflection/scattering, while the second and later NIRS can further extend the coverage in deep NLoS areas, i.e., those areas that are far from the LoS region and barely receive enough signal strength. To achieve this, coordination of multiple NIRS is a key problem to be solved. With more NIRS, the dimension of the optimization problem grows, for which an computationally efficient and effective method to find the global maximum is needed.
\subsection{Joint Communication and Sensing Enhancement with NIRS}
\par In 6G and beyond wireless systems, it is expected that high-level integration of sensing and communication (ISAC) will play an important role. This is even enticing in the THz band, which promises unprecedented millimeter-level sensing accuracy. Even though NIRS is proposed to extend the coverage ability of THz communications, it is also potential to improve the sensing ability. By including NIRS in surrounding environment, the reflection and scattering loss are reduced. This leads that  the back-scattered echo signal for sensing amplifies and therefore, the sensing SNR is increased and higher sensing accuracy can be achieved. However, it is also possible that if being placed in inappropriate positions, scattering from NIRS can cause stronger interference to the sensing echo signals. Experiments are needed to verify whether gain or loss NIRS may bring to THz sensing systems. Furthermore, due to the different metrics of communication and sensing systems, effective algorithms and methods are needed to joint optimize communication and sensing performance, or putting forward a good balance between them.
\section{Conclusion}
In this article, we provided an overview of the non-intelligent reflection surface (NIRS), which is a rough surface simply made of metal materials. The advantages and disadvantages of NIRS compared to IRS are presented. \textit{Still waters run deep} - with almost nil-cost and extremely low fabrication difficulty, NIRS can effectively solve the LoS blockage problem, as well as enhance coverage, channel capacity and even sensing capabilities. Experimental results show that by using the NIRS, the channel capacity in the NLoS area could double on average. To shed light on studying THz NIRS, open problems and future directions are elaborated, including the NIRS channel modeling, reliable design of site-specific usage, deployment and coordination optimization, and joint communication and sensing enhancement. 


\bibliographystyle{IEEEtran}
\bibliography{IEEEabrv,main}
\vfill

\end{document}